\documentclass[twoside]{article}
\usepackage{fleqn,espcrc2}
\input epsf
\input{epsf.sty}

\title{Magnetic relaxation in the "Bragg-glass" phase in BSCCO}

\author{C. J. van der Beek$^{a}$, S. Colson$^{a}$, M. Konczykowski$^{a}$,
M.V. Indenbom$^{a,b}$,  R.J. Drost$^{c}$, and P.H. Kes$^{c}$}

\address{Laboratoire des Solides Irradi\'{e}s, Ecole Polytechnique, 
91128 Palaiseau, France\\ 
\noindent $^{{\rm b}}$Institute of Solid State Physics, 142432
Chernogolovka, Moscow District, Russia\\
\noindent $^{{\rm c}}$Kamerlingh Onnes Laboratorium der Rijksuniversiteit leiden, 
P.O. Box 9506, 2300 RA Leiden, the Netherlands}

\begin{document}

\begin{abstract}
Magnetic relaxation in the Bragg-glass phase of
overdoped Bi$_{2}$Sr$_{2}$CaCu$_{2}$O$_{8}$ crystals 
was investigated using time-resolved magneto-optical visualisation of the 
flux distribution.  This has permitted us to extract the current-voltage 
characteristic, which can be well described by a power-law, although 
fits to a stretched exponential $E \sim \exp( - j_{c} / j )^{\mu}$  with 
$0.3 < \mu < 0.8$ are possible at long times in excess of 100 s.
\vspace{1pc}
\end{abstract}

\maketitle


The mixed state in single crystalline 
Bi$_{2}$Sr$_{2}$CaCu$_{2}$O$_{8}$ (BSCCO) is subdivided in at least 
two other vortex phases: the high temperature (field) vortex liquid, 
separated from the low temperature (field) vortex lattice by a first 
order phase transition (FOT) \cite{Zeldov95II}.  Neutron scattering has shown 
that the low--field phase shows the usual long-range hexagonal vortex 
lattice order \cite{Cubitt93}. Nevertheless, the presence of strong pinning at high
field and low temperature 
\cite{Konczykowski2000} implies 
that disorder must play a role even in this phase, which was 
therefore dubbed ``Bragg glass''\cite{Giamarchi94}. The low pinning 
force density 
in the Bragg glass, and the 
possibility to vary the vortex density over several orders of 
magnitude make this a model system for elastic manifolds in the 
presence of weak random disorder, and offers the 
possibility to check the scaling behavior of displacements and 
displacement correlations. A straightforward means to access this 
behavior is to measure its transport properties, which depends crucially on 
the exponents describing the correlations 
in the random manifold regime \cite{Feigelman89}.

Such measurements at low fields in BSCCO 
are complicated by at least two factors: (i) they is very sensitive to 
the presence of intergrowths and other macro-defects, even in very small concentration 
(ii) the proximity of $H_{c1}$ severely affects the equilibrium vortex distribution and 
flux flow. The latter circumstance renders the 
edge barrier against flux penetration very important; in fact, 
it dominates the magnetization \cite{Chikumoto92II}. In ordinary transport measurements,
all the current is carried by the 
crystal edge, and special measures have to be taken in order to 
measure the bulk properties of the vortex lattice 
\cite{Fuchs98III,Rycroft99}. The latter studies find that in 
optimally doped BSCCO the activation barrier $U$ opposing vortex
creep varies with the current density $j$ as $U \sim j^{-0.5}$ in the range $55 < 
T < 77$ K \cite{Fuchs98III}; in overdoped BSCCO, Rycroft {\em et 
al.} \cite{Rycroft99} find that $U \propto \ln(j_{c}/j)$ for $34 < T < 55$ K, 
reminiscent of two-dimensional vortex dynamics. Still, the high 
currents required and low electric fields limit the use of 
traditional  transport measurements to $T > 30$ K in BSCCO.

The abovementioned problems can be overcome by the 
magneto-optical technique. The  di-

\begin{figure}[h]
    \vspace{-20pt}
    \centerline{\epsfxsize 6.8cm \epsfbox{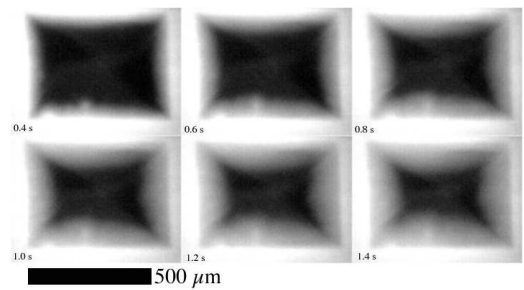}}
    \vspace{-18pt}
\caption{Magneto-optical image of the flux density distribution on 
the  surface of the BSCCO crystal at $T = 18$ K, an applied
magnetic field of 440 G, and successive times $0.4 < t < 1.4 s$. }
\label{fig:Fig1}
\end{figure}

\begin{figure}[t]
    \centerline{\epsfxsize 6.8cm \epsfbox{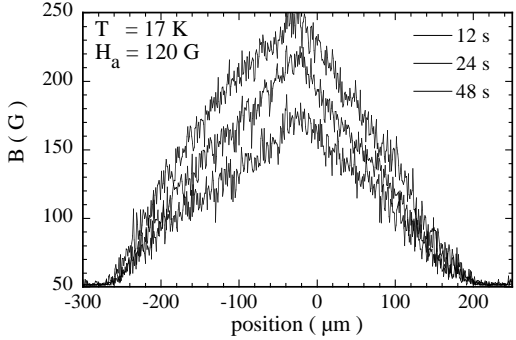}}
    \vspace{-22pt}
\caption{Relaxation of the flux density profile across the surface of 
the BSCCO sample at $ T = 17$ K, 
after field cooling in 520 G and subsequent rapid decrease of $H_{a}$ 
to 120 G. }
    \vspace{-12pt}
\label{fig:Fig2}
\end{figure}

\noindent  rect
visualisation of the flux density on the sample surface 
reveals any macrodefects or other inhomogeneities. From the 
flux density profile, one can easily distinguish between the relative 
contributions of bulk and surface currents. Using this method, we have measured the 
temporal relaxation of flux density profiles in an overdoped BSCCO 
crystal ($T_{c} = 80$ K; size $477 \times 610 \times 20 $ $\mu$m), 
cut and cleaved from a larger crystal grown using the travelling 
solvent floating zone technique. 
The Bragg glass phase in our sample extends to $B_{sp}\sim 600$ G 
at low temperature, much above the first penetration field $H_{p}\sim 100$ G. 
In order to minimize the effects of edge barriers we have 
adopted the following protocol: the field was raised to a value 
$H_{p} \ll H_{a} < B_{sp}$, and then suddenly lowered to the target 
value. The lowering of the field triggers the acquisition of 
magneto-optical images at fixed time intervals; such images are 
displayed in Fig.~1.  Intensity profiles taken across the crystal 
width are converted to flux density profiles (Fig.~2) using a calibration 
measurement above $T_{c}$. The near-perfect linearity of these 
profiles, in correspondence with single--vortex pinning and the Bean 
model, inspired us to take the slope $\partial B / 
\partial x$ as being representative of the bulk screening current 
$j$. Current density vs. time is plotted in Fig.~3.

The result, depicted in Fig.~3(a), shows that $j$ follows a near-perfect power-law behavior, 
$j \propto t^{-0.50}$, for all $14 < T < 23$ K. This corresponds to 
an $I(V)$--law: $E \propto j^{3}$, as found in Ref.~\cite{Rycroft99} 
near the FOT. However, solving Maxwell's equations show that such a
shallow $I(V)$-law cannot result in the sharp linear profiles of 
Fig.~2 \cite{Gilchrist94III}. It is therefore better to parameterize the curves along the 
collective creep prediction: $j \propto [T \ln (t/\tau)]^{1/\mu}$ 
\cite{Feigelman89}. Fig.~3(b) shows that at long times this relation 
is obeyed with a temperature dependent exponent $0.36< \mu < 0.86$, 
as in Ref.~\cite{Fuchs98III}, and close to the theoretical result
of $0.5 < \mu< 0.8$ \cite{Giamarchi94}.


\begin{figure}[t]
\centerline{\epsfxsize 6.7cm \epsfbox{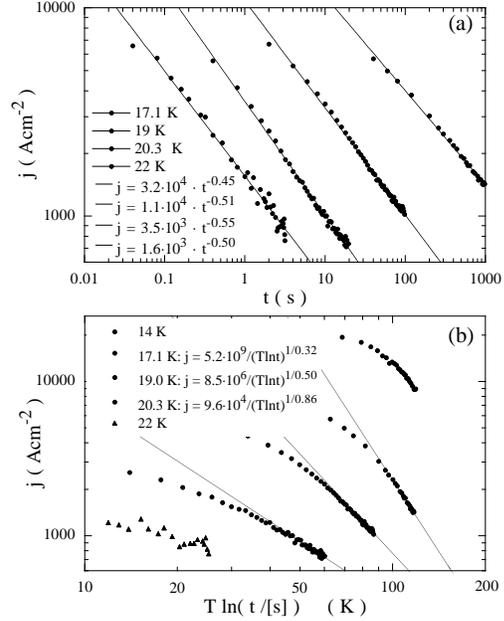}}
\vspace{-18pt}

\caption{(a) Bulk current density vs.time (b) vs. $T\cdot \ln(t)$, for 
different temperatures.}
\label{fig:Fig3}
\vspace{-20pt}
\end{figure}

\end{document}